\documentclass[prl,twocolumn,showpacs]{revtex4}
\usepackage{graphicx}

\newcommand{\cH} {{\cal H}}
\newcommand{\eg} {{\it e.g., }}
\newcommand{\half} {\frac{1}{2}}
\newcommand{\ie} {{\it i.e., }}
\newcommand{\rmd} {{\rm d}}
\newcommand{\rme} {{\rm e}}
\newcommand{\vecq} {{\bf q}}

\begin{document}

\title
{Tension and solute depletion in multilamellar vesicles}

\author{Haim Diamant}

\affiliation{James Franck Institute, The University of Chicago, 
Chicago, Illinois 60637}

\date{November 20, 2001}

\begin{abstract}
We show that a metastable multilamellar vesicle (`onion'),
in contact with excess solvent,
can spontaneously deplete solute molecules from
its interior through an unusual, entropy-driven mechanism.
Fluctuation entropy is gained as the uneven partition of solute 
molecules helps the onion relieve tension in its lamellae.
This mechanism accounts for recent experiments on the
interaction between uncharged
phospholipid onions and dissolved sugars.
\end{abstract}


\maketitle


The richness of self-assembled structures exhibited by amphiphilic
molecules (surfactants) in solution has attracted intensive research
\cite{GompperSchick}. More recently, attention has been turned
toward the behavior of amphiphilic structures under stress
\cite{Cates_book}. An example for such structures are multilamellar 
vesicles (`onions') that form when a 
phase of surfactant bilayers is subjected to shear stress
\cite{Roux}.
Onion structures pose intriguing questions related to
metastability and flow \cite{Roux,Mark,ourEPJ} and have
possible applications, \eg for encapsulation of solid particles
and drug delivery \cite{encapsulation}.
They also appear in biology, \eg as a key ingredient in the
maintenance of the lung surfactant monolayer \cite{lungonion}.

Some properties of onions are rather well understood, \eg the
viscoelasticity of the onion phase \cite{visco}, dependence of
size on shear rate \cite{Roux}, and kinetics of swelling
upon shear-rate change \cite{Leng} or dilution \cite{Mark,ourEPJ}.
Other important features remain unresolved, in
particular, the mechanism of the dynamic lamellar-to-onion
transition \cite{Anton}. Here we address another aspect of onions
that has emerged in recent experiments \cite{Zemb} --- their curious
behavior in the presence of dissolved molecules that do not directly
associate with the bilayers. We show that, as in the case of onion
dilution \cite{ourEPJ}, a key role is played by surface tension
appearing in the lamellae when the onion is under osmotic stress due
to contact with excess solvent. 
The unique way in which tension affects the onion entropy
may lead to a situation where molecules
are depleted from the onion interior for purely entropic reasons.

Typical onions are of micron size, whereas the
inter-membrane spacing is usually of order 10 nm, and the
thickness of an individual membrane is a few nm. Thus,
a typical onion comprises a spherical stack of hundreds to
thousands spaciously packed membranes. Since synthetic bilayers
are usually symmetric and thus do not possess a spontaneous curvature
\cite{Safran_book}, the onion structure is generally {\em not} a
global free energy minimum; the equilibrium state of the membranes, 
depending on water content, is a flat lamellar ($L_\alpha$) phase, a
disordered `sponge' ($L_3$) phase, or a vesicle ($L_2$) phase 
\cite{GompperSchick,Safran_book}.
Yet, the path to this ultimate
state requires onion disintegration or coalescence, involving
breakage of hundreds of membranes per onion, and may take days
\cite{Mark}.
Thus, an onion is in general far from equilibrium with its
environment. Internally, however, the stack may relax on shorter
time scales through the occasional formation of passage defects
between membranes \cite{necks,Leng}. Here we focus on this intermediate
time regime, which may last hours to days, where the onion is in
quasi-equilibrium but cannot annex or release membranes. We ask
how such a structure affects the distribution of small solute
molecules in its vicinity.

In uncharged systems, or when electrostatic interactions are
screened by salt, lamellar phases are stabilized by the Helfrich
interaction \cite{Helfrich_int} --- an entropic, 
fluctuation-induced repulsion between
membranes. For tensionless membranes the free
energy per unit area of this steric effect is
%
$
  f_{\rm st}(d) = {b T^2}/{(\kappa d^2)}
$,
%
where $d$ is the inter-membrane spacing, $T$ the temperature in
energy units, $\kappa$ the bending rigidity of the membranes,
and $b$ a numerical prefactor whose value is under debate; analytic
calculations \cite{Helfrich_int} give $b=3\pi^2/128\simeq 0.2$,
whereas computer simulations \cite{Lipowsky89} find 
$b\simeq 0.06$. The steric repulsion is supplemented by van
der Waals attraction and shorter-range repulsion coming from
hydration effects \cite{membrane_int}. These interactions
determine the equilibrium spacing $d$, as well as the transition
point at which the stack `melts'.

When a lamellar domain coexists in equilibrium with excess
solvent, the inter-membrane pressure equals the external
osmotic pressure. This fact was used in experiments to measure
inter-membrane interactions \cite{Parsegian}. In general, however,
a spherical onion in contact with solvent is out
of thermodynamic equilibrium and an osmotic stress  
may appear. 
Although water molecules can permeate rather freely through the membranes 
on the relevant time scales, further onion swelling is hindered by the 
lack of additional surfactant molecules and the high barriers to membrane 
breakage.
Mechanical equilibrium requires that
any pressure difference between the interior and exterior be balanced 
by surface tension in the lamellae. 
Since tension can be transmitted between membranes through occasional 
passages, 
a tension profile should in general be established throughout
the onion.
Such tension profiles have been shown recently
to stabilize onions against `melting' in an excessively diluted 
environment \cite{ourEPJ}.

The fact that tension has a significant effect on membrane
interactions in onions is demonstrated by a simple argument. 
The parameter that determines the extent of this
effect is the ratio between the inter-membrane spacing $d$ and the
capillary length $l\equiv(2T/\pi\sigma)^{1/2}$, $\sigma$
being the surface tension \cite{RG_tension}. If $x\equiv d/l\ll 1$, 
tension is negligible and the steric interaction is given
by Helfrich's tensionless expression. 
When $x>1$ membrane fluctuations are
strongly suppressed, leading to an exponential decay
of the inter-membrane repulsion \cite{RG_tension}. Suppose, by
negation, that tension were negligible, $x\ll 1$. The interior
pressure then would be roughly $p\simeq 2b'T^2/(\kappa d^3)$,
where the numerical factor $b'$ is a certain fraction of $b$, \ie
of order 0.01--0.1. If there is no external pressure, the interior
pressure is balanced solely by tension, $\sigma=\half Rp \simeq
b'T^2R/(\kappa d^3)$, where $R$ is the onion radius. This gives
$x^2\simeq b'(T/\kappa)(R/d)$. Since typically $R/d\simeq
10^2$--$10^3$ and $\kappa\simeq 1$--$10 T$, one finds $x>1$,
contrary to the initial assumption \cite{ft_binodal}.

We begin the analysis by recalling the effect of tension on the
thermodynamics of a single membrane in a lamellar stack.
The elastic Hamiltonian
governing a single membrane is written in Fourier space as
%
$
  \cH=\half\int\rmd^2 q (\kappa q^4+\sigma q^2) |h_\vecq|^2
$,
%
where $h_\vecq$ is a single mode of height fluctuations. The
fluctuations can be divided, using the Helfrich patch size $\xi$
\cite{Helfrich_int}, into long-wavelength modes, $q<\pi/\xi$, that 
`sense' the confinement due to the adjacent membranes, and 
short-wavelength ones, $q>\pi/\xi$, that are confinement-free. 
The former modes
give rise to a steric repulsion for which, in the case of nonzero
tension, there is no exact expression. A self-consistent
approximation \cite{Seifert} yields
%
$
  f_{\rm st}(d,\sigma) = ({bT^2}/{\kappa d^2})
  (x/\sinh x)^2
$,
%
which adequately coincides with the tensionless expression for small
$x=d/l$ and with renormalization-group calculations
\cite{RG_tension} for large $x$. Since tension suppresses the
steric repulsion, this contribution to the free energy 
decreases with tension. Yet the dominant effect of tension is to
{\em increase} the total free energy by reducing the fluctuation
entropy of individual membranes. Being independent of the spacing
$d$, this contribution, denoted here $f_\sigma$, is usually ignored.
(It is implicitly included in the surfactant chemical potential.)
In the current study, however, we need to consider it explicitly.
It is calculated by tracing $\rme^{-\cH/T}$ over the short-wavelength
modes $|\vecq|>\pi/\xi$. The resulting integral, 
$f_\sigma=\pi T\int_{\pi/\xi}^{\pi/\delta} q\rmd q\ln(1+\sigma/\kappa
q^2)$, 
is dominated by the microscopic cutoff $\delta$, which
is of the order of the membrane thickness. The result is
\begin{equation}
  f_\sigma = \frac{\pi^3T}{2\delta^2} [(1+y)\ln(1+y) - y\ln y],
  \ \ y\equiv\frac{\delta^2\sigma}{\pi^2\kappa}.
\label{fsigma}
\end{equation}

Let us proceed to the thermodynamics of the entire spherical stack.
We employ a phenomenological, `Flory-like' approach
\cite{MilnerRoux,ourEPJ}, where the steric effect is taken into
account accurately whereas the other interactions are incorporated
in a second-virial term. Using the expressions for $f_{\rm st}$ 
and $f_\sigma$ we obtain the free energy density of the onion as
\begin{equation}
  f_{\rm on}(\phi,\sigma) = \frac{bT^2}{\kappa\delta^3}
  \left[ \frac{\phi^3 x^2}{(1-\phi)^2\sinh^2x} -
  \chi\phi^2 \right] + f_\sigma(\sigma)\phi.
\label{fon}
\end{equation}
In Eq. (\ref{fon}) $\phi\equiv\delta/(d+\delta)=N\delta^3/(4\pi R^3/3)$
is the surfactant volume fraction ($N$ being the total number of
surfactant molecules in the onion), and $\chi$ 
characterizes the net attraction balancing the
steric effect. We hereafter consider, for mere
simplicity, uniform concentration and tension profiles within
the onion \cite{ft_uniform}. Thus, the Gibbs free energy of the
entire onion is written as
%
$
  G(T,p_0,N,\sigma) = 
  (4\pi/3) R^3 [f_{\rm on} + p_0] + 4\pi R^2 \sigma
$,
%
where $p_0$ is an external pressure.

We now apply the quasi-equilibrium assumptions mentioned above.
The mechanical equilibrium condition is obtained from the
requirement $(\partial G/\partial R)_{T,p_0,N,\sigma}=0$, 
leading to a Laplace equation, 
$p - p_0 - 2\sigma/R = 0$, with
\begin{equation}
  p(\phi,\sigma) = 
  \frac{bT^2}{\kappa\delta^3} \left[ 2\left(\frac{\phi}{1-\phi}\right)^3
  \frac{x^3\cosh x}{\sinh^3 x} - \chi\phi^2 \right].
\label{pressure}
\end{equation}
In addition, the assumption of internal equilibration implies
a single, uniform chemical potential,
\begin{eqnarray}
  && \mu(\phi,\sigma) = \delta^3
  \left({\partial f_{\rm on}}/{\partial\phi}\right)_{T,p_0,\sigma}
 \nonumber\\
  && = \frac{bT^2}{\kappa} \left[
  \frac{\phi^2 x^2 [(1-\phi)\sinh x + 2x\cosh x]}
  {(1-\phi)^3\sinh^3 x}
  - 2\chi\phi \right]
 \nonumber\\
  && + f_\sigma(\sigma),
\end{eqnarray}
which, as expected, is an increasing function of $\sigma$.

Unlike systems that can only shrink in response to external 
pressure, onions have
the option to release tension instead. For example, if
$N$ and $R$ are constrained so that $\phi$ is fixed, one can still 
increase the pressure, as seen from Eq.~(\ref{pressure}), 
making the tension decrease in response.
%
%
One way to introduce external pressure in practice is to add
large molecules (\eg polymers) that cannot be accommodated
in the inter-membrane spacings and thus
remain in the surrounding solution, exerting an
osmotic pressure on the onion \cite{Deme96,Parsegian}.
%
The current Letter, however, concerns solute
molecules that {\em can} dissolve in the inter-membrane water
layers. In this case the molecules will partition between the
interior and exterior, and the question is whether
their distribution will be even or not. Systems exhibiting
depletion usually involve steric constraints (as for the 
polymers mentioned above) or repulsive interactions. We
consider a case where neither of these is significant; the
molecules can comfortably dissolve in the inter-membrane
layers while being neither attracted to nor repelled from the
membranes. Here we encounter again a delicate consideration of
relaxation times. We assume that permeation of the molecules
through the membranes is kinetically hindered, such that during
the intermediate regime of interest they can be viewed as 
impinging on the membranes and exerting pressure. 
This holds, \eg for sugar molecules, whose diffusion
through an onion stack may take days \cite{Zemb}. One might
wonder then how the distribution of molecules can relax. This
indeed poses a severe experimental difficulty; it has been
overcome, nonetheless, by techniques of freeze drying and
rehydration \cite{Deme96}, where a relaxed distribution equivalent 
to several weeks of molecular diffusion can be achieved.

Let us introduce a dilute reservoir of solute molecules of volume
fraction $\psi_0$. We add to the free energy of the onion
a contribution $G_{\rm s}=(4\pi/3)R^3 f_{\rm s}$,
with
\begin{equation}
  f_{\rm s} = (T/a^3)\left[\psi\ln(\psi/\psi_0)-(\psi-\psi_0)\right],
\end{equation}
where $\psi=Ma^3/(4\pi R^3/3)$ is the volume fraction of the
solute molecules inside the onion ($M$ and $a$ being their total 
number and molecular size, respectively).
Naturally, $G_{\rm s}$ by itself is minimum for an even partition,
$\psi=\psi_0$. Taking the variation of $G_{\rm tot}=G+G_{\rm s}$
with respect to $R$ we get, however, an expected
contribution to the pressure, $p_{\rm s}=(T/a^3)(\psi_0-\psi)$,
which generates a coupling to the onion variables $\sigma$ and
$\phi$ through the Laplace equation. 
We are thus required to find the value of $\psi$ that
minimizes the total free energy $G_{\rm tot}$, given that the
modified Laplace equation,
$p(\phi,\sigma)-p_{\rm s}(\psi)-2\sigma/R=0$, is satisfied.
(For simplicity we take
$p_0=0$, \ie assume that $p_{\rm s}$ is the sole source of osmotic
pressure.)
In order to examine changes in the system (\eg tension release) as
the solute content $\psi$ is progressively increased, an
additional global constraint must be supplied. One simple limit is
to assume that the surfactant volume fraction $\phi$ remains
fixed. In this case the surfactant chemical potential $\mu$ is
bound to decrease with increasing $\psi$ (since $\sigma$ becomes
smaller while $\phi$ is constant). Another simple limit is to
constrain $\mu$. In that case $\phi$ increases (\ie the onion
shrinks) with increasing $\psi$ to compensate for the decrease in
$\sigma$. (In practice, neither of these two constraints is
expected to be strictly accurate; the system may choose another
path over the $\mu(\phi,\sigma)$ surface.)

\begin{figure}[tbh]
\centerline{\resizebox{0.5\textwidth}{!}
{\includegraphics{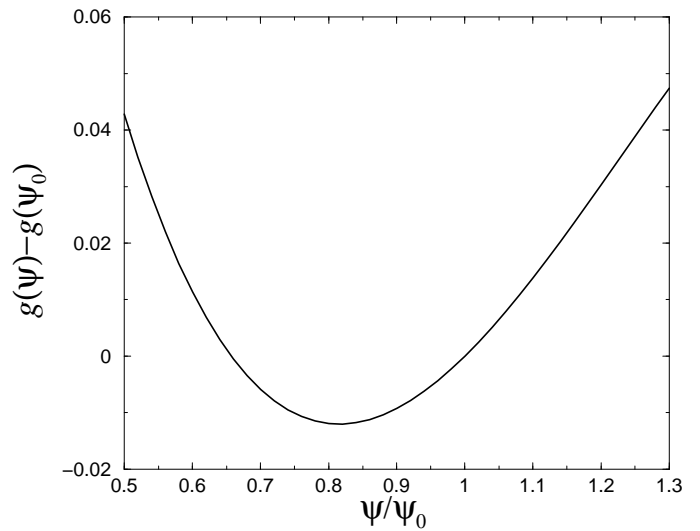}}}
\caption[]{Dependence of free energy on solute
volume fraction. The minimum is obtained for $\psi<\psi_0$,
\ie solute depletion is favorable. 
Parameter values match the conditions
of Ref.~\cite{Zemb}: $\phi=0.59$ (in the absence of solute),
$\kappa/T=10$, $\delta/R=10^{-3}$, $a=\delta$, $b=3\pi^2/128$, 
$\psi_0=0.05$, $\chi=7.57$ (10\% off the binodal; see text).
The free energy per unit volume is given in units of
$2bT^2/(\kappa\delta^3)$.}
\label{fig_min}
\end{figure}

Figure \ref{fig_min} shows the total free energy per unit volume,
$g\equiv G_{\rm tot}/(4\pi R^3/3)$, as a function of $\psi$ for
a given external solute content $\psi_0$, keeping
$\mu$ fixed.
The parameter values ($\phi$ for $\psi_0=0$, $\kappa/T$,
$\delta/R$, $\psi_0$) have been chosen to match the experimental 
system of Ref.~\cite{Zemb}. The value of $\chi$ is taken close
to the binodal point of the stack, $\chi=0.9\chi_{\rm bin}$,
$\chi_{\rm bin}\equiv\phi/(1-\phi)^3$,
mimicking a situation where the bare onion is not so far from
equilibrium coexistence with the surrounding solvent.
As seen in the figure, the favorable solute distribution
is uneven, with a considerable depletion of about 20\%.

\begin{figure}[tbh]
\centerline{\resizebox{0.5\textwidth}{!}
{\includegraphics{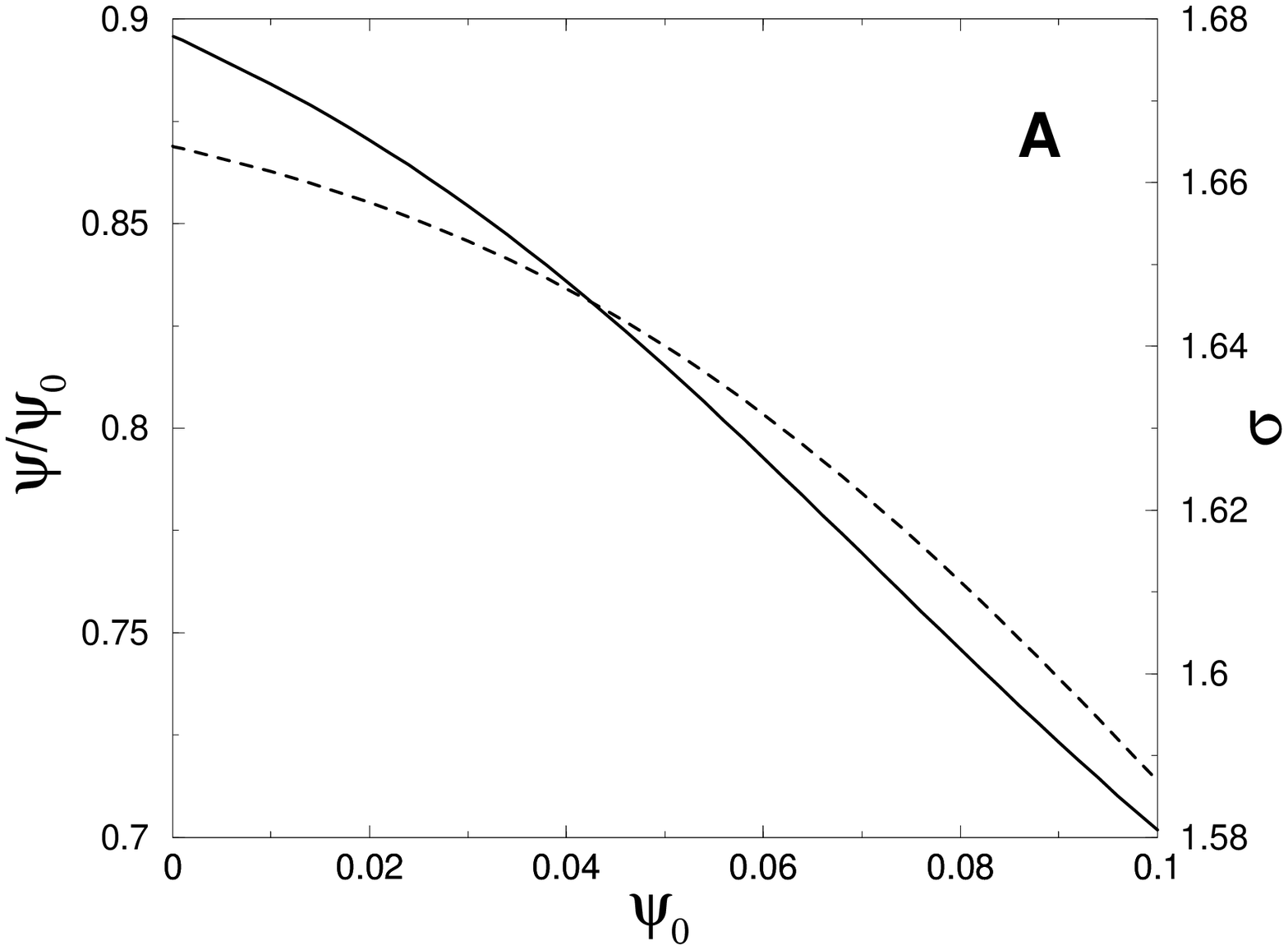}}}
\centerline{\resizebox{0.49\textwidth}{!}
{\includegraphics{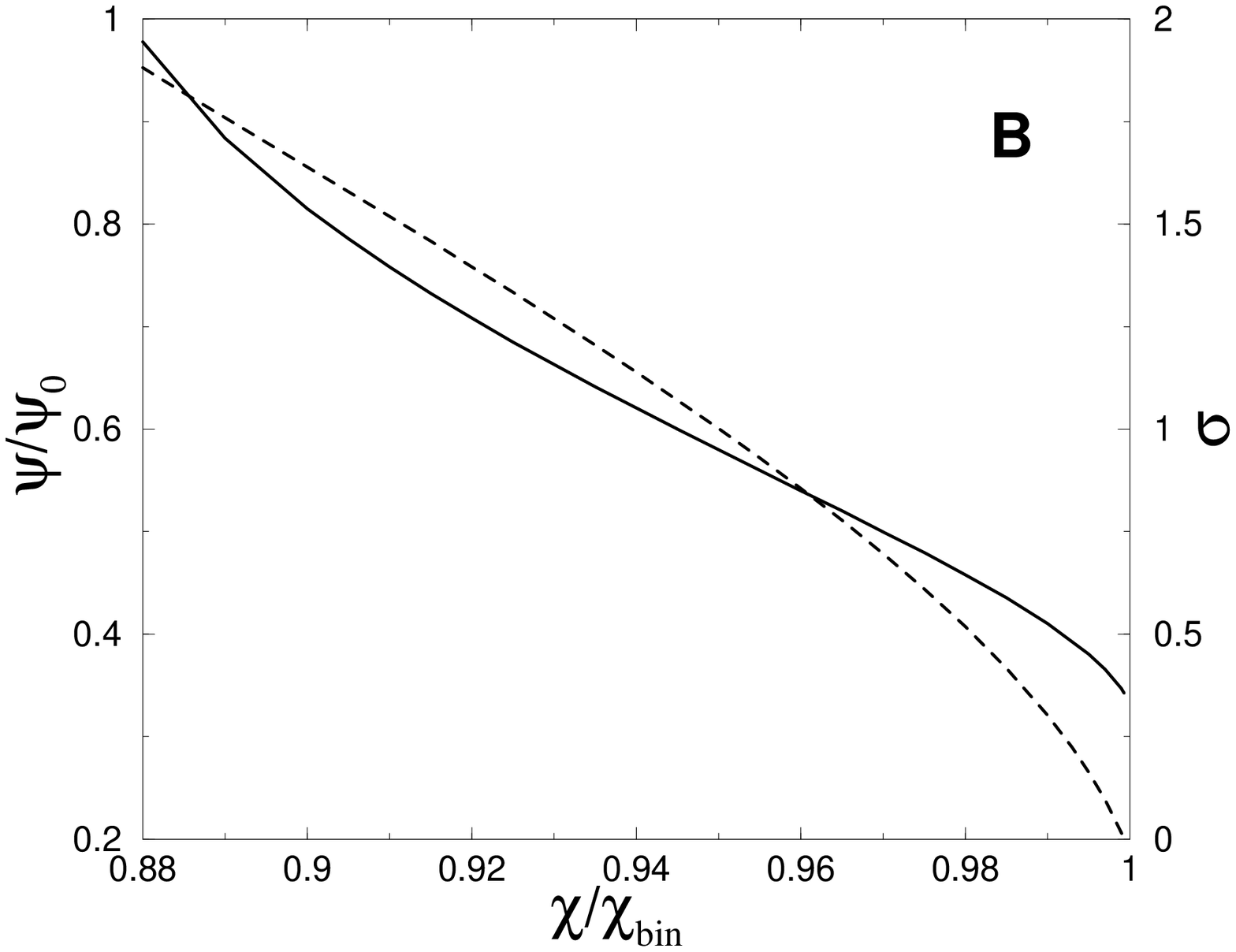}}}
\caption[]{Dependence of depletion (solid curve, left ordinate) 
and tension (dashed curve, right ordinate) on the external 
solute content $\psi_0$ (A) and attraction parameter $\chi$ (B). 
The depletion effect is enhanced as solute molecules are added
and as $\chi$ approaches its binodal value $\chi_{\rm bin}$.
Parameter values are the same as in Fig.~\ref{fig_min}. 
Tension is given in units of $T/\delta^2$.} 
\label{fig_dep}
\end{figure}

The dependence of the favorable $\psi$ and the corresponding
tension release on $\psi_0$ is depicted in
Fig.~\ref{fig_dep}A, demonstrating how the effect
is enhanced as more solute molecules are added.
Finally, we have plotted in Fig.~\ref{fig_dep}B the dependence on
the attraction parameter $\chi$, showing that depletion is in fact
enhanced as one approaches the binodal. The reason is that the
tension in the bare stack becomes lower and the onion is thus more
susceptible to the solute perturbation. This finding implies that
a strong depletion effect can be achieved by slight deviation of
the bare stack from equilibrium coexistence with the excess
solvent. Sufficiently close to the binodal, the bare tension is so
low that the depletion annuls it, and further increase of $\chi$
produces no effect. For the parameter values of Fig.~\ref{fig_dep}B
this occurs extremely close to the binodal.

Intriguing experimental results have been reported recently by
Dem\'e {\it et al}.\ on the interaction of uncharged phopholipid onions
with dissolved hydrocarbonates (small sugars) \cite{Zemb}. 
Using small angle neutron scattering and exploiting the distinct contrast 
match points in the presence and absence of sugar, they determined
that the sugar volume fraction inside the onions was  
tens percent lower than its external value. 
In small angle x-ray scattering the peak corresponding to the lamellar 
periodicity was observed to broaden upon adding sugar, 
indicating an increase in lamellar fluctuations. The authors
attributed the enhanced fluctuations to strong reduction in
bending rigidity due to the added sugar, yet mentioned
that there was no known mechanism for such an effect. Indeed, it
is hard to see how molecules that are well dissolved in the water
layers and do not associate with the membranes could affect their
rigidity. The mechanism suggested here provides a straightforward
explanation for the increased fluctuations, as well as a relation
to the observed depletion --- the enhanced floppiness of the
membranes is caused by tension release rather than reduced
rigidity.

There is an important aspect that is not
captured by the current model --- the onions were found in the experiment
to swell
with increasing sugar content. One possible, previously studied swelling
mechanism invokes the sugar effect on the dielectric constant
of the solvent, which modifies the van der Waals attraction
between membranes \cite{Parsegian}. (This would translate in our
notation to a $\psi$-dependent $\chi$.)
Note that the depletion mechanism presented here is quite general and
should apply, in principle, to other solute molecules such as
dissociated counterions, provided that their permeation through the 
membranes is not too fast. We have considered, however, neutral molecules
whose direct interactions are weak; in the case of charged molecules
strong electrostatic interactions will set in and require a more 
complicated model.

We have treated the interplay between various ingredients of
this complex system on a basic, phenomenological level. 
In particular, we have assumed all quantities (surfactant and 
solute volume fractions, membrane tension) to be uniform 
throughout the onion which, in fact, cannot be strictly correct
\cite{ft_uniform}. The next natural step would be to consider
nonuniform profiles and see, \eg whether the solute forms a
nontrivial profile inside the onion. 
It would be interesting to check also whether the unusual
depletion mechanism suggested here is exploited in biological 
systems, such as the lamellar bodies in the lung.

\begin{acknowledgments}
I am grateful to B.\ Dem\'e and Th.\ Zemb for sharing 
unpublished results,
and to M.\ Cates for helpful comments. 
This work was supported by the National Science Foundation
under Awards Nos.\ DMR 9975533 and 0094569, its 
MRSEC program under Award No.\ 9808595,
and the American Lung Association (RG-085-N).
\end{acknowledgments}



\begin{thebibliography}{99}
\bibitem{ft_binodal}
If the inter-membrane spacing is exactly equal 
to the binodal (coexistence) value, there will be no pressure difference
and hence no tension. Yet, a deviation of
merely 5\% from this value is sufficient to yield $x\simeq 1$.
\bibitem{ft_uniform}
Note that this assumption is in principle self-inconsistent. A
nonzero, uniform tension profile must be accompanied by a
nonuniform pressure profile, leading to a nonuniform concentration
profile \cite{ourEPJ}. Here we employ the uniform simplification
merely for brevity, assuming that the actual variations
are moderate on the scale of the onion size and thus should not
affect the qualitative results.
\end{thebibliography}
\end{document}